\documentclass[twocolumn,showpacs,preprintnumbers,amsmath,amssymb,prc]{revtex4}

\usepackage{graphicx}
\usepackage{dcolumn}

\begin{document}

\title{New variational Monte Carlo method with an energy variance extrapolation\\
for large-scale shell-model calculations}

\author{Takahiro Mizusaki$^{1}$ and  Noritaka Shimizu$^{2}$ }
\affiliation{
$^{1}$ Institute of Natural Sciences, Senshu University, Tokyo 101-8425, Japan \\
$^{2}$ Center for Nuclear Study, University of Tokyo, Hongo, Tokyo 113-0033, Japan \\
}

%\date{\today}

\begin{abstract}
We propose a new variational Monte Carlo (VMC) method with an energy variance extrapolation
for large-scale shell-model calculations.
This variational Monte Carlo is a stochastic optimization method with
a projected correlated condensed pair state as a trial wave function, and is 
formulated with the $M$-scheme representation of projection operators, the Pfaffian and 
the Markov-chain Monte Carlo (MCMC).
Using this method, we can stochastically calculate approximated yrast energies and 
electro-magnetic transition strengths. 
Furthermore, by combining this VMC method with energy variance extrapolation,
we can estimate {\it exact} shell-model energies.   

\end{abstract}


\maketitle

Shell-model calculations have been a central issue in studies of nuclear structure, 
and constant effort has been made to solve large-scale shell-model problems. 
Exact diagonalization, which is a standard method, has recently been able to handle large-scale problems with $O(10^{10})$ dimension
\cite{caurier,shimizu-okinawa}. However, 
computational feasibility in this method is limited and strongly depends on the size of
single particle space and valence nucleon numbers.
To overcome this problem and to extend the feasibility of shell-model calculations, 
various methods\cite{smmc,dmrg,vampir,CCM,qmcd,mcsm,exlanc,papen,exmcsm} with different kinds of algorithms have been proposed and improved.

In this paper, we propose a new method for shell-model calculations
using the Markov-chain Monte Carlo (MCMC).
This new method is a variational Monte Carlo (VMC)
with a projected correlated condensed pair state as a trial wave function. 
This new VMC can stochastically give not only energies but also electro-magnetic 
transition strengths, although these values are approximated.
To estimate the {\it exact} shell-model energies over the limitations of 
variational formulation, we use the energy variance extrapolation, 
which has been studied in Refs.\cite{imada1,sorella} and has been successfully applied to the 
nuclear shell model \cite{exlanc,exmcsm}.
As this extrapolation needs a series of systematically approximated wave functions, 
we introduce a truncation scheme based on the spherical basis in the form of
a projection operator.

In this study, projection operators are implemented by using the Monte Carlo in a novel way.
In fact, particle number projection, magnetic quantum number projection, parity projection, 
and projection onto truncation spaces can be implicitly performed without numerical integrations, 
and angular momentum projection is performed by two-dimensional numerical integration.

Here we consider a new variational formulation of shell-model calculations. 
As a trial wave function for a nuclei with $N_\pi$ valence protons and $N_\nu$ 
valence neutrons, {\it i.e.,} $N = N_\pi + N_\nu$, 
we take $| \psi \rangle$ as  
\begin{equation}
|\psi\rangle=GP|\phi\rangle
\label{trial}
\end{equation}
and
\begin{equation}
| \phi \rangle = 
\left( \sum f_{kk^{\prime}}c_{k}^{\dagger}c_{k^{\prime}}^{\dagger}\right)^{N/2}|0\rangle
\end{equation}
where $f$ is a skew-symmetric matrix. The $|0\rangle$ is an inert core and the $c_i^{\dagger}$'s are 
proton or neutron creation operators ($i=1,\cdots, N_\pi$ : proton and $i=N_\pi+1,\cdots, N$ : neutron).
By this parametrization, proton-neutron pairing correlation is included, in addition to the proton-proton 
and neutron-neutron pairing correlations. 
For simplicity, to present this formulation, we assume that the components of $f$ are real numbers. 
We can also use a complex number although this would require some modifications.
The $P$ is a projection operator and we use the following two kinds.
One is 
\begin{equation}
P=P^{I_0} P^{\pi} P_{M} 
\label{Proj_m}
\end{equation}
where $P^{I_0}$, $P^{\pi}$ and $P_M$ are projectors of the $z$-component of isospin $I_0$, parity $\pi$ and $z$-component of angular momentum $M$, respectively.
The other is 
\begin{equation}
P=P^{I_0} P^{\pi} P^J 
\label{Proj_j}
\end{equation}
where $P^J$ is a projection onto angular momentum $J$.
The $G$ is a correlation factor as
\begin{equation}
G=e^{-\sum_{i\leq j} \alpha_{ij} n_i n_j }
\label{Gfactor}
\end{equation} 
where  $\alpha$'s are variational parameters.
The $n_i$ is taken as a number operator for orbit $i$, 
so that the $G$ is commutable with the angular-momentum projector.
The $G$  can be easily evaluated as will be discussed later.
%because we take a diagonal form in the $M$-scheme representation 

For the trial wave function in Eq.(\ref{trial}), the energy expectation value is 
given by 
\begin{equation}
E=\langle H\rangle =\frac{\langle\phi\left|PGHGP\right|\phi\rangle}{\langle\phi\left|PGGP\right|\phi\rangle}.
\label{penergy}
\end{equation}
To evaluate this, we introduce an alternative representation of the projection operator 
in Eq.(\ref{Proj_m}),
which can be shown by
the so-called $M$-scheme states defined as
\begin{equation}
\left|m\rangle=c_{m_{1}}^{\dagger}c_{m_{2}}^{\dagger}\right.\cdots c_{m_{N}}^{\dagger}| 0\rangle,
\label{mscheme}
\end{equation}
where $m=(m_{1},m_{2},\cdots m_{N})$. 
As the $M$-scheme state is an eigenstate of the 
$z$-components of the isospin and angular momentum, and parity,
the projection operator is given by 
\begin{equation}
  P^{I_0} P^{\pi} P_M  = \sum_{m\in\{M^\pi \}} |m\rangle \langle m|
\label{proj_mscheme}
\end{equation}
where the summation is restricted within the $M$-scheme states with given quantum numbers,
that is, particle numbers $N_\pi$ and $N_\nu$, and magnetic and parity quantum numbers $M$ and  $\pi$.
By this representation, $\langle m|P|\phi\rangle$ can easily be calculated as
\begin{equation}
\langle m| P^{I_0} P^{\pi} P_M |\phi\rangle=\langle m|\phi\rangle\delta_{z(m),M^{\pi}}
\label{mphi}
\end{equation}
where $z(m)$ gives the parity and magnetic quantum numbers of the $M$-scheme state $|m\rangle$.

By introducing the $M$-scheme representation of the projection operator,
the $G$-factor becomes a c-number because the $G$-factor is diagonal in the $M$-scheme, that is, 
\begin{equation}
G|m\rangle=G(m)|m\rangle, 
\label{Gdiag}
\end{equation}
where the $G(m)$ is an expectation value of the $G$ concerning the $M$-scheme state $|m\rangle$.
Moreover the matrix elements of the $H$ between $M$-scheme states can also be evaluated, and
we denote them as
\begin{equation}
  \langle m|H|m' \rangle = h_{m,m'}
\label{Hope}
\end{equation}
where $|m \rangle$ and $|m' \rangle$ have the same quantum numbers and $h$ is generally very sparse
because a shell-model Hamiltonian consists of one-body and two-body interactions.

By these relations (\ref{Gdiag}) and (\ref{Hope}), Eq.(\ref{penergy}) can be rewritten as
\begin{equation}
E=\sum_{m\in\{M^\pi \}}\rho(m)E_{L}(m),
\label{mc_energy}
\end{equation}
where the local energy $E_{L}(m)$ is defined as 
\begin{equation}
E_{L}(m)=\sum_{m^{\prime}\in\{M^\pi   \}} h_{m,m'}  \frac{\langle m^{\prime}|P|\phi\rangle G(m^{\prime})}
{\langle m|P|\phi\rangle G(m)},
\label{localenergy}
\end{equation}
and the sampling density $\rho(m)$ is defined by
\begin{equation}
\rho(m)=\frac{|\langle m|P|\phi\rangle G(m)|^{2}}
{\sum_{m\in\{M^\pi  \}}|\langle m|P|\phi\rangle G(m)|^{2}},
\label{rho}
\end{equation}
where $\rho(m)\geq 0$ and $\sum_{m\in\{M^\pi \}}\rho(m)=1$.

In numerical calculations, the energy formula Eq.(\ref{mc_energy}) is not practical 
because the dimension of the $M$-scheme space becomes intractably huge as the size of single particle
space and proton and neutron numbers increase. 
Eq.(\ref{mc_energy}), however, is suited to Monte Carlo calculations because
if we can generate a set of the $M$-scheme states $|m\rangle$ that obey the occurrence ratio $\rho(m)$ 
by the Monte Carlo sampling,
the energy expectation value can be estimated by
\begin{equation}
E\sim \frac{1}{N_0}\sum_i E_{L}(m_{i}),
\label{mcsum}
\end{equation}
where the $N_0$ is a number of Monte Carlo samples.

In this formulation, the projection operator appears only in the projected overlap 
between the $|\phi\rangle$ and the $M$-scheme state $|m\rangle$.
In the case of the projection operator Eq.(\ref{Proj_m}), 
the projected overlap simply becomes the overlap $\langle m|\phi\rangle$ as shown in
Eq.(\ref{mphi}), and this overlap can be given by the Pfaffian as
\begin{equation}
\langle m|\phi\rangle=\left(\frac{N}{2}\right)!{\rm Pf}(X_{m})
\end{equation}
where 
$(X_{m})_{ij}\equiv f_{m_{i}m_{j}}-f_{m_{j}m_{i}}$\cite{Tahara}. 
The definition of the Pfaffian is shown in the Appendix.

Next we delve into the Markov-chain Monte Carlo.
To evaluate Eq.(\ref{mcsum}) by the Monte Carlo method, 
a set of {$|m\rangle$} whose distribution obeys Eq.(\ref{rho}) is needed.
Here we consider a random walker $|m\rangle$ on the $M$-scheme space with the given
proton and neutron numbers, parity, and magnetic quantum numbers.
A random walker  $|m\rangle$ moves to $|m'\rangle$ on the $M$-scheme space
with the same quantum numbers in the following way:
\begin{itemize}
\item We choose two nucleons  $(m_i,m_j)$ in the $|m\rangle$ randomly and annihilate the nucleons
in the $|m\rangle$. We call the resultant state $|m_I\rangle = c_{m_j}c_{m_i}|m\rangle$.
\item We sum up the magnetic quantum numbers, the $z$-component of the isospin and the parity of the chosen 
nucleons.  
\item In $|m_I \rangle$, we randomly choose two available unoccupied states, 
$(m'_i,m'_j)$, whose summed quantum numbers are the same as those of $(m_i, m_j)$.
Then, we create two nucleons on chosen states.  
We call the resultant state $|m'\rangle = c^\dagger_{m'_i} c^\dagger_{m'_j}|m_I\rangle $.
\end{itemize} 

Transition of random walker can be controlled in two ways.
One is the Metropolis-Hasting (MH) algorithm.
Whether or not a random walker $|m\rangle$ moves to  $|m'\rangle$ 
depends on the ratio $p(m')$ as 
\begin{equation}
p(m')=\left|\frac{\langle m'|P|\phi\rangle G(m')}
        {\langle m|P|\phi\rangle G(m)}\right|^{2}.
\end{equation}
If $p(m')\geq 1$, the walker $|m\rangle$ always moves to $|m'\rangle$. 
If $p(m')<1$, according to the $p(m')$, we determine whether or not the walker $|m\rangle$ moves 
to $|m'\rangle$.

The other is the Gibbs sampling algorithm.
When we consider a random walker $|m\rangle$, 
  we choose a removed pair randomly and obtain $|m_I\rangle$.
  Then, we calculate all $p(m')$'s for possible
  $|m'\rangle = c^\dagger_{m'_i}c^\dagger_{m'_j} |m_I\rangle$
  conserving appropriate quantum numbers. 
  Finally, we choose $|m'\rangle$ by transition probability
  $W(m_I\rightarrow m')$ defined by
  \begin{equation}
    W(m_I \rightarrow m')=\frac{p(m')}{\sum_{m''} p(m'')}.
  \end{equation}   
These algorithms satisfy detailed balance and ergodicity.  

Next we consider how to optimize the parameters' $\alpha$'s and $f$'s 
in the trial wave function.
To use the steepest descent method, a gradient vector 
needs to be evaluated by the Monte Carlo method. 
Each component of the gradient vector is given by straightforward
calculations as
\begin{equation}
\frac{\partial E}{\partial \alpha_{ij}}= 2\langle HO_{\alpha_{ij}}\rangle-
2\langle H\rangle\langle O_{\alpha_{ij}}\rangle
\end{equation}  
where 
\begin{equation}
O_{\alpha_{ij}}=n_i n_j
\end{equation} 
and
\begin{equation}
\frac{\partial E}{\partial f_{ij}}= 2\langle HO_{f_{ij}}\rangle-
2\langle H\rangle\langle O_{f_{ij}}\rangle
\end{equation}  
where $O_{f_{ij}}$  is an operator, the matrix elements of which are as follows
\begin{equation}
O_{f_{ij}}(m)=\frac{1}{\langle m|P|\phi \rangle}\frac{\partial}{\partial f_{ij}}\langle m|P|\phi \rangle.
\end{equation} 
Sophisticated derivation is shown in Ref.\cite{Tahara}. The derivation needs some extensions
if we use complex numbers for $f_{ij}$.
The gradient vector obtained by the Monte Carlo method suffers from stochastic noises.
To reduce such noises we use the stochastic reconfiguration (SR) method \cite{sorella_sr}, the
details of which are shown also in Ref.\cite{Tahara}.
In this way, we can numerically evaluate the gradient vector and can optimize the parameters of 
the present wave function based on the steepest descent method.

In this formulation, proton and neutron number projections, magnetic quantum number projection,
and parity projection
can be implicitly performed without numerical integration, 
while angular momentum projection can not.
The $J$-projection $P^{J}_M$ \cite{ringschuck}  is given by  
\begin{equation}
P^{J}_M\equiv\sum_{K}g_{K}P_{MK}^{J},
\label{jproj}
\end{equation}
and
\begin{equation}
P_{MK}^{J} \equiv\frac{\bar J}{8\pi^{2}}\int d\Omega D_{MK}^{*J}(\Omega)R(\Omega)
\end{equation}
where the $\Omega$ stands for Euler's  angles $(\alpha,\beta,\gamma)$
and $\bar J =2J+1$.
The $R(\Omega)$ is a rotational operator 
and $D_{MK}^{J}(\Omega)$ is Wigner's $D$-function.
The $g$'s are additional variational parameters. 

The angular momentum operator is rewritten by
\begin{equation}
P_{MK}^{J}=P_{M} \tilde{P}_{MK}^{J}
\end{equation}
where 
\begin{equation}
\tilde{P}_{MK}^{J}\equiv\frac{\bar J}{4\pi}\int d\beta d\gamma \sin\beta d_{MK}^{J}(\beta)e^{-iK\gamma}
e^{iJ_y\beta}e^{iJ_z \gamma}.
\end{equation}
The $d_{MK}^{J}(\beta)$ is Wigner's $d$-function.
The projected overlap $\langle m|P^{I_0} P^{\pi} P^J_M |\phi\rangle$ becomes   
\begin{equation}
\langle m|P^{I_0} P^{\pi} P^J_M |\phi\rangle
=\sum_K g_K\langle m| \tilde{P}_{MK}^{J}|\phi\rangle\delta_{z(m),M^{\pi}},
\end{equation}
with $P^J_M = \sum_K g_K P^J_{M,K}$.
The projected overlap concerning $ \tilde{P}_{MK}^{J}$ is
\begin{equation}
\langle m| \tilde{P}_{MK}^{J}|\phi\rangle=
\frac{\bar J}{4\pi}\int d\beta d\gamma \sin\beta d_{MK}^{J}(\beta)e^{-iK\gamma}R(\beta,\gamma)
\end{equation}
with
\begin{equation}
R(\beta,\gamma)=\langle m\left| e^{iJ_y\beta}e^{iJ_z \gamma} \right|\phi\rangle,
\end{equation}
which can be calculated by the Pfaffian.
Thus, the angular momentum projection can be performed by the two-dimensional integration,
which is a distinguishable feature in this formulation.  

Finally, we consider how to compute electro-magnetic transition strengths. 
To evaluate them, angular momentum projection is indispensable and 
unnormalized initial and final states can be denoted by
\begin{equation}
|\psi_\sigma; J_\sigma, M_\sigma \rangle=
G\sum_{K_\sigma} g_{K_\sigma} P^{J_\sigma}_{M_\sigma,K_\sigma}|\phi_\sigma\rangle
\label{jwave}
\end{equation}
where $\sigma=i,f$. The $J_i$ and $J_f$ are
spins of the initial and final states and the parameters of these wave functions 
are $f$'s, $\alpha$'s and $g$'s.

The B(E2; $J_i \rightarrow J_f$), for instance, is defined as
\begin{equation}
B(E2; J_i \rightarrow J_f)=\frac{1}{2J_i+1} 
\left| 
\langle \tilde{\psi} _f || Q ||  \tilde{\psi}_i \rangle
\right|^2
\end{equation}
where $Q$ is a quadrupole operator, and $| \tilde{\psi}_i \rangle$ and $| \tilde{\psi}_f \rangle$ are 
normalized initial and final wave functions with spins $J_i$ and $J_f$, respectively.
For this Monte Carlo evaluation, noting the following relation as
\begin{equation}
\left| 
  \langle \tilde{\psi}_f|Q|\tilde{\psi}_i\rangle
\right|^2
= \frac{\langle \psi_f|Q|\psi_i\rangle}{\langle \psi_f|\psi_f\rangle }
\frac{\langle \psi_i|Q|\psi_f\rangle}{\langle \psi_i|\psi_i\rangle },
\end{equation}  
where $|\tilde{\psi}_\sigma \rangle=|\psi_\sigma \rangle / 
\sqrt{\langle \psi_\sigma|\psi_\sigma \rangle} $,
we can perform the MCMC  for $\frac{\langle \psi_i|Q|\psi_f\rangle}{\langle \psi_i|\psi_i\rangle }$ and
$ \frac{\langle \psi_f|Q|\psi_i\rangle}{\langle \psi_f|\psi_f\rangle }$, respectively.
For the MCMC of 
the $ \frac{\langle \psi_f|Q|\psi_i\rangle}{\langle \psi_f|\psi_f\rangle }$ term with
Eq.(\ref{jwave}),
the sampling density $\rho(m)$ is given by
\begin{equation}
\rho(m)=\frac{|\langle m|P^{J_f}_{M_f}|\phi_f\rangle G(m)|^{2}}
{\sum_{m\in\{M_f^{\pi_f}\}}|\langle m|P^{J_f}_{M_f}|\phi_f\rangle G(m)|^{2}},
\end{equation}
and local quadrupole strength $Q_L(m)$ is given by
\begin{equation}
Q_{L}(m)=\sum_{m^{\prime}\in\{M_i^{\pi_i}\}}\langle m|Q|m^{\prime}\rangle 
\frac{\langle m^{\prime}|P^{J_i}_{M_i}|\phi_i\rangle G(m^{\prime})}
{\langle m|P^{J_f}_{M_f}|\phi_f\rangle G(m)}
\end{equation}
where $P^{J_\sigma}_{M_\sigma} = \sum_{K_\sigma} g_{K_\sigma} P^{J_\sigma}_{M_\sigma, K_\sigma}$.
Note that we take different $M$-scheme spaces for $|m\rangle$ ($J_z|m\rangle=M_f|m\rangle$) and $|m'\rangle$ ($J_z|m'\rangle=M_i|m'\rangle$). 
Thus, the strength of the electro-magnetic transition can be evaluated in the MCMC.
 
Next, we numerically investigate the present variational Monte Carlo method.
Here we consider the $^{56}$Ni in the $pf$ shell with the 
GXPF1A interaction\cite{Gxpf1a}, of which dimensions are about 1.09 billion in the $M$-scheme.  

\begin{figure}[h]
  \includegraphics[width=8cm]{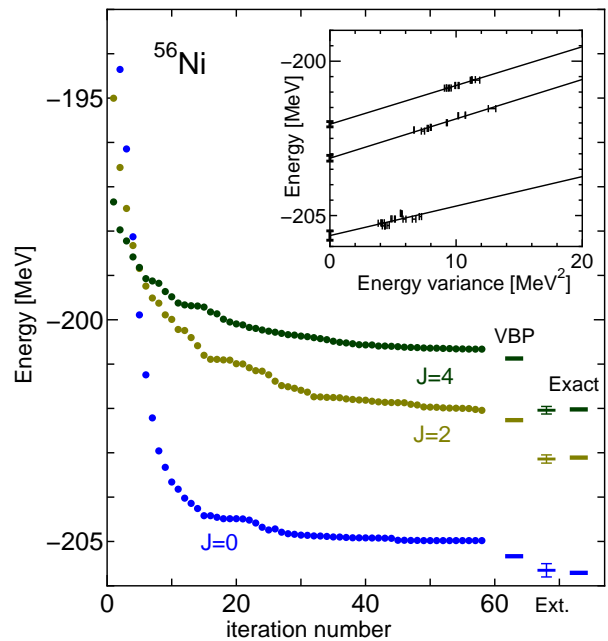}
  \caption{(Color online) 
    Convergence patterns of energies with $J_{z}=$ 0, 2 and 4 states
    as functions of the iteration number for $^{56}$Ni in the $pf$ shell with the GXPF1A
    \protect\cite{Gxpf1a} interaction. 
    The $J$-projected energies (VBP) onto the converged states at $J_{z}=J$ space 
    are shown with the extrapolated (Ext.) and exact (Exact) shell model energies.   
    In the inset, extrapolations of energies with $J$ = 0, 2 and 4
    as functions of energy variance are shown. 
  }
  \label{fig1}
\end{figure}

First, we carry out the VMC with a $J_{z}=J$ space where we consider the state 
with angular momentum $J$. 
To perform the VMC, we prepare an initial wave function randomly and
simulate the sampling density  $\rho(m)$ in Eq.(\ref{rho}) by the Monte Carlo. 
Here we use the Gibbs sampling algorithm.
After appropriate burn-in steps ($\sim 1000$), a random walker moves more than 5000 steps 
in the $M$-scheme space. These numbers depend on the acceptance ratio and required accuracy of numerical calculations.
In this way, we prepare several tens random walkers and 
estimate the energy and energy gradient vector with statistical errors.
With the aid of the SR technique, we modify all the 
variational parameters of the wave function and repeat this optimization process 
until the energy variation goes to zero.
In Fig. \ref{fig1}, we show the convergence patterns of the energies  
with  $J_{z}=$ 0, 2  and 4 states as functions of the iteration number for $^{56}$Ni. 
The statistical error during the optimization procedure
is a few tens keV, which is too small to be shown in this figure.

%===============  fig. 1  ========================================

Next, we stochastically calculate the $J$-projected energy
by carrying out a $J$-projection on the converged wave functions
at $J_{z}=J$ space.
We call this stochastic VBP (variation-before-projection).
Note that the parameters of the wave function are optimized 
concerning number, parity, and $J_z$ projected energy.
Here, we take $g_K = \delta_{KJ}$ for simplicity 
because we carry out the $J$-projection onto the wave functions optimized 
in the space with $J_z=J$.
For $J=0$ state, after 1000 steps as a burn-in, a random walker moves 500000 steps in 
the $M$-scheme space. We evaluate the energy by 10 random walkers.
The energy is -205.333 $\pm$ 0.004 MeV for $^{56}$Ni.
We show the $J$-projected energies for $J=0$, 2, and 4 with the label VBP in Fig. \ref{fig1}. 
The $J$-projection improves while there are still sizable differences 
between these VBP energies and exact shell model energies.
In this formulation, we can stochastically evaluate electro-magnetic 
transition strengths with statistical errors.
The calculated B(E2)'s  with the VBP wave functions for $0\rightarrow 2$ and  $2\rightarrow 4$ are  
690 $\pm$ 8 and 134 $\pm$ 2 $e^2fm^4$, respectively.
Here we use effective charges $e_{\pi}=1.5$ and $e_{\nu}=0.5$.

To overcome the variational limitation of the VMC, 
we introduce the energy variance 
extrapolation, which is a technique to estimate the exact energy from a series of approximated 
wave functions in a well-controlled way.
This technique is based on a well-defined scaling property for energy eigenvalues.
We define a difference $\delta E$ between energy eigenvalue 
$\langle H \rangle$ in a given subspace and 
exact energy eigenvalue $\langle H \rangle _0$, that is,
$
\delta E = \langle H \rangle - \langle H \rangle_0,
$
and an energy variance  $\Delta E$ in the subspace is also defined as
$
\Delta E={\left\langle H^2 \right\rangle -\left\langle H \right\rangle ^2}.
$
The difference $\delta E$ vanishes linearly or quadratically 
as a function of the energy variance $\Delta E$ \cite{exlanc}.
By this scaling property, we can estimate the exact shell-model energies as the limit of zero energy variance. 

To apply this technique to the VMC, we introduce a truncation scheme in the form of
a projection operator. In the case of the $pf$ shell, due to a relatively large gap 
of spherical single particle energy between the $f_{7/2}$ orbit and others  $r$ 
($f_{5/2}$, $p_{3/2}$ and $p_{1/2}$ orbits),
particle-hole excitations across this shell gap form
truncation spaces, $\oplus_{s\le t}(f_{7/2})^{A-40-s}(r)^s$. 
The $P^t$ is
a projection onto this truncation space and is added to the projection in Eqs.(\ref{Proj_m},
\ref{Proj_j}).
This projection operator is quite simply realized by the restriction of the summation in Eq.(\ref{proj_mscheme}) and we can easily include this truncation scheme in the present VMC. 

It is noteworthy that, in the Monte Carlo calculations, we find a fast computational 
method of the energy variance within the truncated space $t$ as 
\begin{equation}
\left\langle H^2 \right\rangle_t \sim \frac{W_t}{N_0}\sum_{i\in(t+2)} (E_{L}(m_{i}))^2,
\label{mcsumv}
\end{equation}
where $E_{L}$ is the local energy defined in Eq.(\ref{localenergy}),
$N_0$ is a number of Monte Carlo sampler, and $W_t$ is a reweighing factor
defined by the ratio between the number of random walkers within the $(t+2)$-space,
that is, $N_0$,   
and one of random walkers within the $t$-space.  By this sampling method, we can 
avoid the explicit calculation of four-body interaction for energy variance
and its computation becomes possible.  

In the inset of Fig. \ref{fig1}, 
energies of the  VMC calculations with different truncation spaces and different initial 
conditions are plotted as functions of the energy variance,  $\Delta E$.
To the limit of zero energy variance,
we can linearly extrapolate the energies with statistical errors obtained by the $\chi^2$ fitting.
The extrapolated energies agree with the exact ones. 
In the same way, we can extrapolate the B(E2). 

In conclusion, we have proposed a new variational Monte Carlo method with energy variance 
extrapolation for large-scale shell-model calculations.
We have presented a formulation of wave function optimization based on the MCMC.
In this method we can calculate approximated energy, other matrix elements, and electro-magnetic transitions for yrast states.
Combining the VMC with energy variance extrapolation, we can estimate exact shell-model energies.
This is an alternative extension of the VMC and is free from the sign-problem.
By taking $^{56}$Ni in the $pf$ shell, we have shown the feasibility of large-scale shell-model calculations.  Note that the present calculations can be carried out with a single core of the common PC. For larger computations, parallel computation plays a significant role.

For further improvement of the present method, the stochastic VAP
(variation-after-projection) concerning $J$-projection can improve
accuracy of energy and energy variance 
to enhance the reliability of energy variance extrapolation. 
We are pursuing the stochastic VAP including its parallel computation as especially fitted 
to a state-of-the-art massive parallel computer, the results of which will be presented elsewhere.  

For energy variance extrapolation, we introduced the particle-hole truncation scheme into the VMC
while we can use the arbitrary basis-truncation scheme. The excitation-energy truncation scheme
in the No-core Shell Model \cite{Nocore} is also promising. We will investigate this direction 
in the future.

Finally, this study is motivated by a study of the Hubbard model\cite{Tahara}. 
The present study sheds light onto a new way of projection in these 
stochastic calculations, an aspect of which may also be useful in applications of condensed matter physics.    
 
One of the authors (N.S.) was supported by Grants-in-Aid for
Young Scientists (20740127) from JSPS and the HPCI Strategic Program 
from MEXT.
 
\hspace{5cm}
\appendix{Appendix}

The Pfaffian is defined for a $2n\times 2n$ skew-symmetric matrix $A$ as   
\begin{equation}
Pf(A)\displaystyle \equiv\frac{1}{2^{n}n!}\sum_{\sigma\in S_{2n}}
{\rm sgn}(\sigma)\prod_{i=1}^{n}a_{\sigma(2i-1)\sigma(2i)}
\end{equation}
where $\sigma$ is a permutation of $\{1,2,3,\cdots , 2n\}$, ${\rm sgn}(\sigma)$ is its sign,
$S_{2n}$ is symmetry group and $a$'s are matrix elements of $A$.
An efficient computation of the Pfaffian can be seen {\it e.g.} in Ref.\cite{phaff}.

%-----------------------------------------------------------------------

\end{document}